\documentstyle[prl,twocolumn,aps]{revtex}
\tighten
\begin{document}
\draft

\title{Noise suppression due to long-range Coulomb interaction: \\
Crossover between diffusive and ballistic transport regimes}
\author{T. Gonz\'alez, O. M. Bulashenko,\cite{byline} J. Mateos, and D. Pardo}
\address{
Departamento de F\'{\i}sica Aplicada, Universidad de Salamanca,
Plaza de la Merced s/n, \\ E-37008 Salamanca, Spain}
\author{L. Reggiani}
\address{
Istituto Nazionale di Fisica della Materia, Dipartimento di Scienza
dei Materiali, \\ Universit\`a di Lecce, Via Arnesano, 73100 Lecce, Italy}
\author{J. M. Rub\'{\i}}
\address{
Departament de F\'{\i}sica Fonamental, Universitat de Barcelona,
Av. Diagonal 647, \\ E-08028 Barcelona, Spain}

\twocolumn[
\date{10 February 1997) \\
(cond-mat/9703123, 
Semicond. Science and Technol., v.12, 1053 (1997)}

\maketitle

\widetext \vspace*{-0.5in}

\begin{abstract}
\begin{center} 
\parbox{14cm}{
We present a Monte Carlo analysis of shot-noise suppression due to
long-range Coulomb interaction in semiconductor samples under a 
crossover between diffusive and ballistic transport regimes.
By varying the mean time between collisions we find that the strong 
suppression observed under the ballistic regime persists under 
quasi-ballistic conditions, before being washed out when a complete 
diffusive regime is reached.
}\end{center} 
\end{abstract}
]\narrowtext

It has been recognized that shot noise and thermal noise are not different
sources of noise, but just special limits of a more general concept (see
discussion by Landauer
\cite{landauer}), which can be treated in a unified 
way \cite{landauer,lino}.
This point has been clarified by studying noise properties of
mesoscopic systems such as quantum point contacts, quantum wires, quantum
dots, etc \cite{jong} in which both ballistic ($\lambda_p\gg L$) and
diffusive ($\lambda_p\ll L$) regimes of carrier transport are accessible
(here $\lambda_p$ is the mean free path and $L$ is the device length).

A matter of particular interest is the suppression of shot noise 
resulting from the correlation among carriers in their transmission 
through a device (see some recent experiments 
\cite{li,liefrink,reznikov,kumar,steinbach} and the
status of theoretical developments \cite{jong}).
Among the different mechanisms of noise suppression most attention has
been paid to the correlations imposed by the Pauli principle (Fermi
statistics) under degenerate conditions.
In this case, the reduction of noise has been found for both ballistic
\cite{lesovik} and diffusive regimes \cite{beenakker}, and the crossover
between them
has been studied theoretically by de Jong and Beenakker \cite{jong95}.

However, as emphasized by Landauer \cite{landauer}, apart
from Pauli correlations, Coulomb interaction between carriers may
also be a source of noise suppression.
By using an ensemble Monte Carlo simulator self-consistently coupled with a
Poisson solver (PS) we have calculated the shot-noise 
suppression associated with long-range Coulomb interaction in the case 
of perfect ballistic transport \cite{shot1}. Here we address the case of 
low-density carrier concentration typical of semiconductors, thus
avoiding any effect due to statistical degeneracy. Under this condition
we have found that this suppression is stronger as space-charge effects
become more important, and can achieve higher levels for increasing
values of the parameter $\lambda=L/L_{Dc}$, where
$L_{Dc}=\sqrt{\varepsilon k_BT/q^2n_c}$ is the Debye length corresponding
to the carrier concentration at the injecting contact ($T$ is the lattice
temperature, $k_B$ is Boltzmann constant, $\varepsilon$ is the dielectric 
constant, $q$ is the electron charge, and $n_c$ is the carrier concentration 
at the contact).
Therefore, the longer the device length, the larger the shot-noise
suppression is expected to be, provided that transport remains
ballistic.
However, with increasing the device length (or lattice temperature) 
carrier transport actually becomes diffusive and the disappearance of
noise suppression associated with Coulomb correlation is expected. 
Note that in the case of metals all charge fluctuations 
are screened out and Coulomb correlations play no role 
in the dc transport regime.
It is the aim of the present letter to investigate the behavior of noise
suppression due to the long-range Coulomb interaction among carriers 
in the crossover region from ballistic to diffusive transport regimes
in nondegenerate semiconductors.
We will show that noise suppression still remains important under 
quasi-ballistic conditions, before being washed out by the 
action of scattering mechanisms.

For the calculations we consider the following simple
model: a lightly doped active
region of a semiconductor sample sandwiched between two heavily doped contacts
(of the same semiconductor) injecting carriers into the active region.
The contacts are considered to be ohmic (the voltage drop inside them
is negligible) and they remain always in thermal equilibrium.
Thus electrons are emitted from the contacts according to a thermal-equilibrium
Maxwell-Boltzmann distribution and they move inside the active region, following
the semiclassical equations of motion, by undergoing elastic and isotropic
scattering.
The fluctuating injection rate at the contacts, which is associated
with the diffusion current of a homojunction, is taken to follow a 
Poissonian statistics. 
Accordingly, the time between two consecutive electron injections
is generated with a probability per unit time $P(t)=\Gamma e^{-\Gamma t}$,
where $\Gamma=\frac{1}{2}n_c v_{\rm th}S$ is the injection rate,
$v_{\rm th}=\sqrt{2k_B T/\pi m}$ is the thermal velocity, 
$S$ is the cross-sectional area of the device and
$m$ is the electron effective mass.
The simulation is one-dimensional in real space
(Poisson's equation is solved only in one dimension) and three-dimensional
in momentum space.

For the simulations we have used the following set of parameters:
$T_0=300\,K$, $m=0.25\,m_0$ ($m_0$ being the free electron mass),
relative dielectric constant $\varepsilon$=11.7, $L=2000\,\AA$ and 
$n_c=4\times 10^{17} {\rm cm}^{-3}$, much higher than the sample doping
(here taken of $10^{11} {\rm cm}^{-3}$, but the same results are obtained
up to $10^{15} {\rm cm}^{-3}$).
The above set of values yields for the dimensionless parameter
$\lambda=L/L_{Dc}$, which characterizes the importance
of the electrostatic screening, the value $\lambda=30.9$.
Along with $\lambda$, we introduce the dimensionless 
ballistic parameter $\ell=v_{\rm th}\tau/L$ which characterizes the crossover 
between diffusive and ballistic transport regimes.
The average time between collisions in the bulk $\tau$ is assumed to be
independent of energy \cite{remark}, and it is varied from
$10^{-15} {\rm s}$ to $10^{-11} {\rm s}$, so that both regimes
of carrier transport, ballistic ($\ell\gg 1$) and diffusive
($\ell\ll 1$), are covered. For the PS, a time step
of 2 fs and 100 meshes in real space are taken.

We apply a $dc$ voltage and calculate the time-averaged
current $I$ and the current autocorrelation function $C_I(t)$
by means of the ensemble Monte Carlo simulator self-consistently
coupled with the PS \cite{JAP}.
We stress that in our approach the number of electrons $N$ inside 
the sample is not required to be fixed. Carriers are injected at $x=0$
and $x=L$ inside the active region of the device according to $P(t)$.
When a carrier exits through any of the contacts it is cancelled from 
the simulation statistics, which accounts only for the carriers which are
inside the device at the given time $t$.
Thus, $N(t)$ fluctuates in time due to the random injection from the contacts
and we can evaluate both the time-averaged value $\langle N \rangle$
and its fluctuations $\delta N(t)=N(t)-\langle N \rangle$.
To illustrate the importance of the effects associated with the 
long-range Coulomb interaction,
we provide the results for two different simulation schemes.
The first one involves a {\em dynamic} PS, so that any fluctuation of
space charge arising because of the random injection from the contacts 
causes a redistribution of the potential which is self-consistently updated 
by solving the Poisson equation at each time step during the simulation. 
The second scheme makes use of a {\em static} PS, so that only 
the stationary potential profile is calculated;
once the steady state is reached, the PS is switched off and
carriers move in the {\em frozen} non-fluctuating electric field profile.
{\em Both schemes give exactly the same average current and steady-state 
spatial distributions of all the quantities, but the noise 
characteristics are quite different.}

By varying the ballistic parameter $\ell$ through $\tau$ 
we find that the steady-state
spatial profile of the potential does not change significantly.
It always exhibits a minimum near the cathode due to the presence of
space charge whenever the current does not saturate.
This minimum acts as a potential barrier for the electrons moving between the
contacts, so that those electrons which do not have 
enough energy to pass over the barrier are reflected back to the contacts.
The most important fact is that the barrier height and, as a consequence,
the transmission through it depends on the current, which is crucial in
calculating the noise characteristics.

Fig.\ \ref{sup} presents the results for the low-frequency spectral
density of current fluctuations $S_I=2\int_{-\infty}^{\infty}C_I(t)dt$
as a function of $\ell$ for an applied voltage of $U=40k_B T/q$.
For the present value of $\lambda$ this voltage corresponds to
the maximum suppression of noise, reaching a value of 0.045.
The results are normalized to $2qI_s$,
where $I_s=q\Gamma=\frac{1}{2}qn_c v_{\rm th}S$ is the saturation current
(maximum current that our contacts in thermal equilibrium can provide).
Both static and dynamic PS cases are shown.
The curve $2qI$, reported for comparison, illustrates how the average
current changes with the ballistic parameter $\ell$.
In the diffusive regime ($\ell\to 0$) the current decreases linearly with 
$\ell$ (or, which is the same, with $\tau$) since scattering processes
prevent carriers from gaining velocity in the field direction \cite{remark2}.
The noise in this regime corresponds to thermal (nonequilibrium) noise,
and thus $S_I$ is also proportional to $\ell$.
Both PS schemes provide identical results, which means that the action
of the scattering mechanisms prevails over that of the self-consistent
field fluctuations. 
When approaching the ballistic regime, the current increases sublinearly
with $\ell$ until it saturates at the value 0.65$I_s$ in the 
ballistic limit ($\ell\to\infty$). 
Starting from $\ell=0.01$ the noise departs from the thermal
behavior. In the static case it obeys the classical formula $2qI$ 
typical
of full shot noise (Poissonian statistics, no correlation among
carriers is taken into account).
On the contrary, in the dynamic case $S_I$ is 
strongly reduced with respect to $2qI$ as $\ell$ increases due to the
correlations associated with the self-consistent potential.
A suppression factor of more than one order of magnitude (0.045) is obtained 
in the ballistic limit. 
In the intermediate quasi-ballistic region the dynamic noise 
is observed to exhibit a maximum value at $\ell\approx 0.1$.
Then, the noise spectral density decreases with increasing $\ell$,
which is somewhat surprising in view of a corresponding increase
of the current.
Notice that the difference between the two schemes starts to appear
already at $\ell\sim 0.01$, which means that the long-range Coulomb
interactions influence the noise even for almost ``diffusive''
regimes when an electron undergoes $\sim 10^2$ scattering events
while crossing the active region of the device.
Another remarkable fact is that $S_I$
in the static case already follows the $2qI$ law for $\ell=0.03$, 
although under such conditions a carrier undergoes about 30 scattering
processes in its transfer between contacts.

To clarify the role played by different fluctuating mechanisms in 
giving the total noise we decompose the current autocorrelation 
function into three contributions $C_I(t)=C_V(t)+C_N(t)+C_{VN}(t)$,
respectively given by

\begin{mathletters}\label{con}
\begin{eqnarray}
C_V(t)&=&\frac{q^2}{L^2} \langle N\rangle^2 \langle \delta v(t') \delta
v(t'+t)\rangle \\
C_N(t)&=&\frac{q^2}{L^2} \langle v\rangle^2 \langle \delta N(t') \delta
N(t'+t)\rangle \\
C_{VN}(t)&=&\frac{q^2}{L^2} \langle v\rangle\langle N\rangle \langle \delta
v(t')
\delta N(t'+t) \nonumber \\
&&\phantom{xxxxxx} + \delta N(t') \delta v(t'+t)\rangle
\end{eqnarray}\end{mathletters}
in the above equations $C_V$ is associated with fluctuations in the
mean carrier velocity $\delta v(t)=v(t)- \langle v\rangle$,
$C_N$ with fluctuations in the carrier number
and $C_{VN}$ with their cross-correlation
 \cite{JAP}.
The spectral densities of the current fluctuations corresponding to these
contributions are illustrated in Fig.\ \ref{decomp}.
For $\ell\to 0$, i.e. in the thermal noise limit,
$S_N\to 0$, $S_{VN}\to 0$ and $S_I\to S_V$, which means that
velocity fluctuations are the main source of noise.
Here $S_V$ is proportional to $\tau$, thus corresponding to a
diffusive behavior.
It should be noted that the difference in the velocity-fluctuation terms
$S_V$ for the two schemes is practically negligible
in the whole range of values 
taken by the ballistic parameter $\ell$ [Fig.\ \ref{decomp}(a)].
On the contrary, for the other two contributions the difference becomes 
dramatic starting from $\ell\approx 0.03$. 
The number fluctuations increase with the ballistic character of 
transport. 
However such an increase is much more pronounced in the static case 
than in the dynamic case where, under the
action of the self-consistent potential fluctuations, number fluctuations 
are significantly suppressed.
The velocity-number correlations, represented by $S_{VN}$, 
show an opposite behavior. 
Their contribution is positive in the static case, while it is 
negative in the dynamic case.
Furthermore, for the current spectral densities calculated in the dynamic 
scheme, $S_N$ and $S_{VN}$ are of opposite sign thus practically
compensating each other. 
As a consequence, the current noise is considerably suppressed in the 
dynamic case with respect to the static case.
This result reflects the fact that as carriers move 
ballistically (or quasi-ballistically) through the active
region, the dynamic fluctuations of the electric field modulate 
the transmission
through the potential minimum and smooth out the current fluctuations 
(of shot-noise type) imposed by the random injection at the contacts. 
The presence of scattering mechanisms, by randomizing the carrier velocity, 
reduces the fluctuations associated with carrier number
before washing them out completely when the carrier motion is fully diffusive, 
leading the current noise to become thermal.

In conclusion, within a Monte Carlo scheme 
we have investigated the influence of long-range Coulomb
interaction on shot-noise suppression in the transition region from 
ballistic to diffusive transport regimes.
We have found that the strong noise suppression observed under ballistic
conditions remains important also in a wide quasi-ballistic region, 
before being completely washed out in the diffusive regime.
The role played by number and velocity-number fluctuations in the total 
noise has been quantitatively estimated and found to explain the 
microscopic origin of the suppression mechanism.

This work has been partially supported by the Comi\-si\'on Interministerial de
Ciencia y Tecnolog\'{\i}a through the project TIC95-0652.

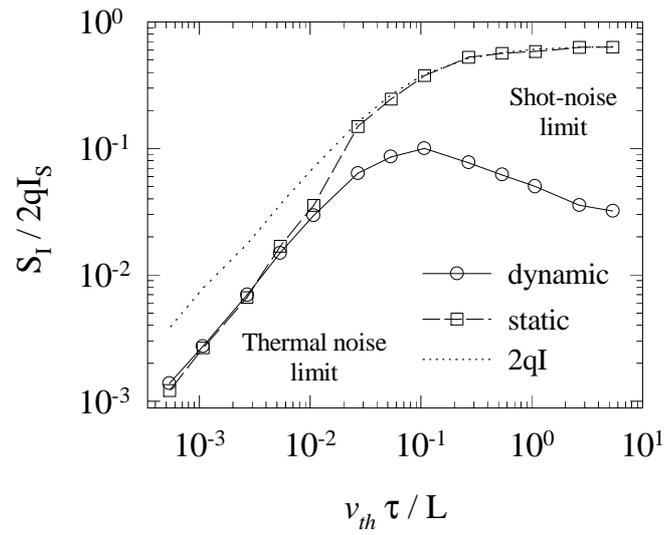
\begin{figure}
\caption{
Current-noise spectral density $S_I$ vs the ballistic parameter 
$\ell=v_{\rm th}\tau/L$
calculated by using static (squares, dashed line) and dynamic 
(circles, solid line) potentials.
The dotted line represents $2qI$.
}\label{sup}\end{figure}

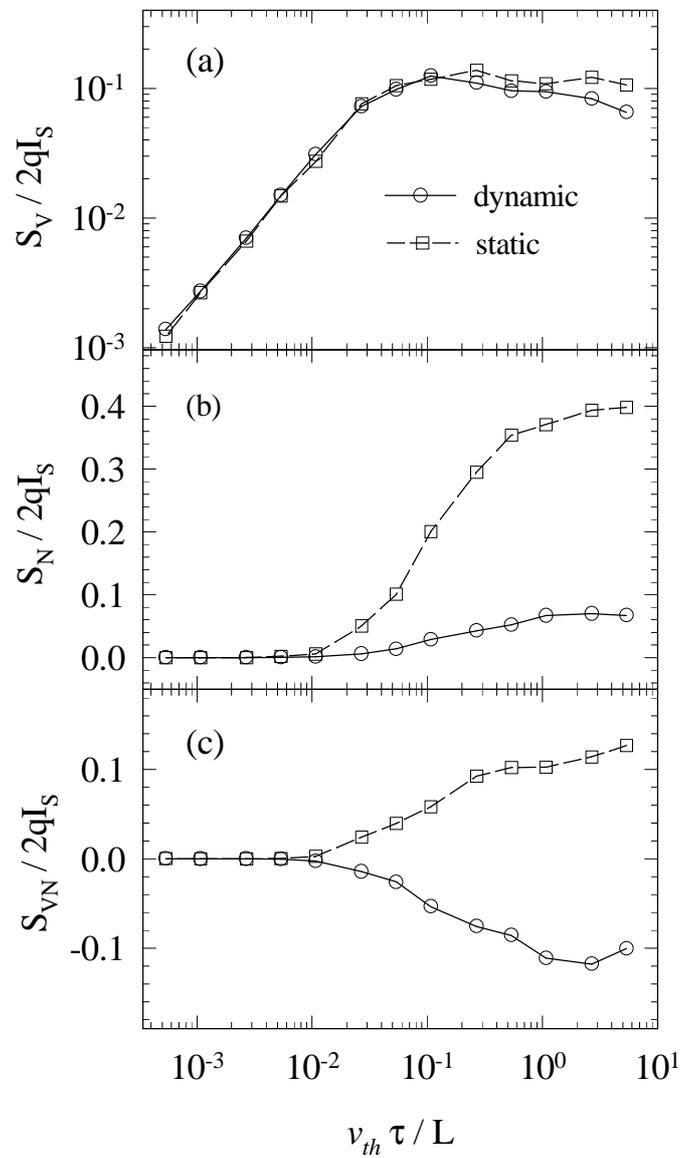
\begin{figure}
\caption{
Decomposition of the $S_I$ of Fig.\ 1
into: (a) velocity, (b) number, and (c) velocity-number contributions vs
ballistic parameter $\ell=v_{\rm th}\tau/L$ for the static 
(squares, dashed line) and dynamic (circles, solid line) potentials.
}\label{decomp}\end{figure}


\begin{references}
\bibitem{remark}
We assume elastic isotropic scattering and use an energy-independent $\tau$
for simplicity, but our model can be easily extended to include any inelastic
collision with energy-dependent relaxation time.
\bibitem{remark2}
The linear dependence of the current on $\ell$ in the diffusive regime
is a consequence of our model. The choice of an isotropic, elastic scattering 
with energy-independent $\tau$ leads to ohmic response of the carriers
(constant mobility $\mu=q\tau/m$) at any voltage when transport is dominated
by collisions.
\end{references}
\end{document}